# Effect of Social Media on Website Popularity:

## Differences between Public and Private Universities in Indonesia

Hanum Putri Permatasari
Department of Information System
Gunadarma University
Depok, Indonesia

Silvia Harlena
Department of Information System
Gunadarma University
Depok, Indonesia

Donny Erlangga
Department of Informatics Management
Gunadarma University
Depok, Indonesia

Reza Chandra
Department of Information System
Gunadarma University
Depok, Indonesia

---

Abstract— Social media has become something that is important to enhance social networking and sharing of information through the website. Social media have not only changed social networking, they provide a valuable tool for social organization, activism, political, healthcare and even academic relations in the university. The researchers conducted present study with objectives to a). examine the academic use of social media by universities, b). measure the popularity and visibility of social media owned by universities. This study was delimited to the universities in Indonesia. The population of the study consisted both on public and private universities. The sample size comprised totally of 264 universities that their ranks included both in Webometrics and 4ICU in July 2012 edition. The social media which was examined included Facebook, Twitter, Flicker, LinkedIn, Youtube, Wikipedia, Blogs, social network community owned by the university and Open Course Ware. This study used an approach for data collection and measurement: by using Alexa and Majestic SEO. Data analysis using the Pearson Chi-square for social media ownership that using data ordinal and independent t test for examining effects of social media on website popularity. The study revealed that majority of the social media users used Facebook, then followed by Twitter. There are also most significant differences for result of popularity by Alexa Rank and visibility by Majestic SEO in universities whether used social media or no.

---

## I. INTRODUCTION

There is an incredible increase in use of World Wide Web and related technologies. The number of internet users in the world about 2.4 billion and the growth in number of internet users are 566.4% in the years between 2000 and 2012 [1]. In parallel to penetration in use of internet, new trends have been emerged, as in scientific issues. Scientist, scientific groups and institutions have benefited from internet as a way of scientific communication and dissemination. Internet also became a global library including billions of scholar and non-scholar publications. Rapid growth requires those dealing with quantitative data to develop a new web-based measurement.

In 2012 Internet users in Indonesia has increased from the previous year, from 42 million to 55 million. This figure means that 23% of the total population of Indonesia. Usage is dominated by large cities and only 4.1% use in rural areas.

These numbers are expected to rise significantly in the years to come as technology becomes more affordable. The number of people who use mobile devices reached 29 million people. This means that over 50% of internet users in Indonesia use mobile devices to surf the Internet [2].

In the internet age, university websites are very important for to their stakeholders and there is a need to assess their ranking. In the recent years, ranking systems have gained wide attention in many universities all around the world. Several popular organizations produce worldwide university rankings, including Webometrics, 4ICU, ARWU, Times Higher Education World University Rankings and QS World University Rankings.

A university has a website to introduce their campus, and the relevant agencies, resources and services, students, alumni, and many so on. One of important factor for the success of the university is its web accessibility and visibility. [3].

Social media has become something that is important to enhance social networking and sharing of information through the website. The aim is to promote even increase web visibility and activity. Social media have not only changed social networking, they provide a valuable tool for social organization, activism, political, healthcare and even academic relations in the university. The popularity and increase use of social media in everyday life is not a surprise.

Social media is Internet based technology which promotes opportunities to social interaction; among its users. It is enhanced through new communication tools and sites that are called; social networking sites. Internet-based tools and audio-visual technology with the ability to retrieve, store, connect and take the features that make the authors publish their work, including through blogs and receive comments on it [4]. Wikis has the ability to promote and facilitate the creation of a common through academic collaboration [5], Social bookmarking is an online catalog of hyperlinks that help users who want to share [6], Facebook, Twitter, and LinkedIn, including the social networking site called SNS that has the ability of online promotion [7]

University started using social media as a component of the overall marketing mix. This study aims to explore the use of social media in college academically. The purpose of this study was (a) examine the use of social media by the academic colleges and (b) measure the popularity and visibility of social media owned by universities.

## II. THEORETICAL BACKGROUND

### A. Social Media in Higher Education

Social media has a major impact on higher education by creating a virtual learning environment to increase distributed learning [8]. Learners can shape communities and interact with each other in cyberspace. They can exchange learning experiences, research and academic opportunities. Several reasons justify the use of social media, including web 2.0 for academic purposes. The comprehensive report 2008, [9] showed that students have taken advantage of social media to support teaching and learning, sharing and communication among learners. With the increasing use of social media by academicians and learners, it seems suitable to keep in mind the prediction of Armstrong & Franklin (p.27) that "Universities will lose their privileged role as a primary producer of knowledge, and gatekeeper to it, as knowledge becomes more widely accessible through other sources and is produced by more people in more ways" [10] in its true sense. Many researchers have addressed different areas of using social media at various academic and social levels. The available literature on social networking media put forward useful ideas for implementing in higher education [11].

It mostly stressed on creating contents with focusing on the way of sharing, interacting and collaborating and socializing through social media. Social media is used for various aims in higher education for intensifying the study experiences of learners by providing them with students support services, including e-mentoring, e-feedback and other e-facilities [12].

Social media is being used to extend communication among and between learners and their communities. The Facebook was suggested as a means of communication for interacting with students [13]. Professional use of social media seems significant for almost all professions but setting up novice teachers for overcoming with internet generation/learners having an information age mindset and for working in increasingly digital educational environments is of greater importance [14].

Social media is represented to be communication facilitator and learners of the day wish their higher education institutions using social networking for supporting classroom work [15]. They take effort for using social networking sites for academic interactions [16] In supplemental, such networking may connect the learning gap informally between "digital native" students and "digital immigrant" faculty [17]

Following [18]; [19], the digital divide is defined as the gap between the students who have access to digital technology at home and those who do not. The factors causing the gap are socioeconomic status, ethnicity and geographic location. The phenomenon of the digital divide has become one of the most popular topics for many researchers and policy-makers since the late 1990s, throughout the countries around the world to experience it to some extent. Even the most developed nations, the United States, for example, face the problem of the digital divide [20] [21].

### B. Related Research

Some challenges and issues associated with the use of social media and networking sites. These challenges and issues reported by higher education students [22] and social media policy makers in higher education institutions [23], including the moral and social concern. The study of Cain, Scott, & Akers [24] confirmed the use of Facebook by students of pharmacy with low understanding of issues relating to e-professionalism and accountability, and same with the findings by Olson et al. on on pre-service teacher education students.

Conflicts are common occur among faculty and students in the use of social media and networking sites as one-third of students are not pleasant that teachers should look at Facebook at all [25] due to some moral issues. One is the use of social media to provide commentary and open response to a message or post. The qualitative study conducted by [26] on the students of United Kingdoms University using Facebook, it was reported that they used Facebook to criticize a learning experience, exchange information about their program (s) related matters, extending moral support to one another and, paradoxically, promote themselves to become an academic not involved or incompetent. The following table shows related research that relevant with this study (Table 1).

TABLE I.    AVERAGE ALEXA RANK (HAVE/HAVEN'T SOCIAL MEDIA)

| Researcher | Type of Socmed | Samples | Methodology |
|---|---|---|---|
| J. Armstrong et al. (2008) | Web 2.0 | 180 (academic & administrative support from 5 countries) | Qualitative survey |
| Chuck Martin (2009) | All social media | 6 HE | Qualitative survey |
| M. D. Roblyer et al. (2010) | Facebook | 182 (faculty & student) | Pearson Chi square |
| I. Ahmed et al. (2011) | All social media | 6 HE | Case Study |
| Patient Rambe (2011) | Facebook | 165 students | Critical ethnography |
| I. Hussain et al. (2012) | All social media | 600 students | Case Study |
| Nikleia Eteokleous et al. (2012) | Facebook | 5 HE | Qualitative survey |

### III.    METHODOLOGY

The study was conducted with the main focus on evaluation of using social media to increase visibility and popularity of the university official websites. The population of the study consisted of 264 universities in Indonesia. There are 53 public and 211 private universities which their ranks included both in Webometrics and 4ICU in July 2012 edition. The social media which was examined included Facebook, Twitter, Flicker, LinkedIn, Youtube, Wikipeda, Blogs, social network community owned by the university and Open Course Ware. This study used an approach for data collection and measurement: by using Alexa and Majestic SEO because they offer special function that search for matches only in web elements such as referring domain, external backlinks, citation flow, trust flow, global rank, id rank and reputation link. Collection was conducted on December 26-29 2012.

Data analysis focused on analysis of the digital divide from two perspectives, types of universities in Indonesia: public and private university and also locations of university: that in Java and outside Java. Proof of the digital divide using the Pearson chi-square for social media ownership that using data ordinal and independent t test for examining effects of social media on website popularity.

### IV.    RESULT AND DISCUSSION

The population of the study consisted of 264 universities in Indonesia. There are 53 public and 211 private universities which their ranks included both in Webometrics and 4ICU in July 2012 edition. From 53 public universities, there are 24 universities which located in Java and 29 out Java. While, there are 153 universities which located in java and 58 out Java from 211 private universities totally (Fig. 1)

Java is main island and center of Indonesia. Education in Indonesia mostly concentrated in the major cities of Java. In

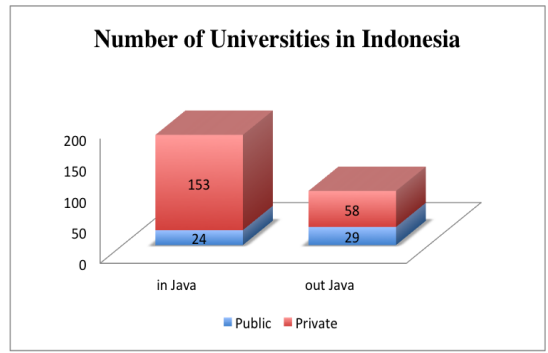

Figure 1.    Number of Universities in Indonesia

Java, also relatively provides the high speed internet access than out Java. Number of lecturers in public universities are 67.939 and 43.875 lecturers in private. Number of students in public universities are 574.800 and 931.266 students in private (Fig. 2).

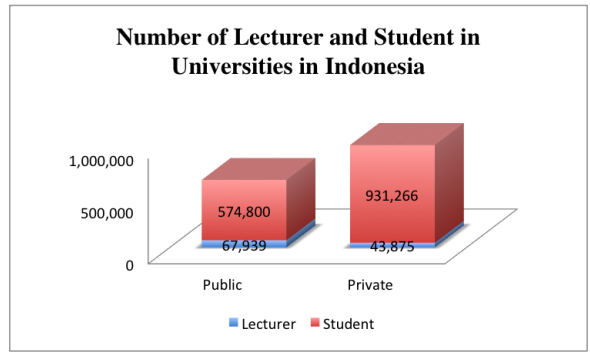

Figure 2.    Number of Lecturer and Student in Universities in Indonesia

Fig. 3 reflects various types of social media which universities usually use. According to the figure, 26% universities were using Facebook which 28% were public, while 72% private; and 25% Twitter (25% public and 75% private), whereas used 3% Flickr (50% public and 50% private), 14% LinkedIn (43% public and 57% private), 10% Youtube (37% public and 63% private), 77% Wikipedia (26% public and 63% private), 1% OCW (67% public and 33% private), and had 30% their own students blog (35% public and 65% private) also had 38% their own staff blog (34% public and 66% private). Some universities also have special own community or social network (68% public and 32% private) and community blog (36% public and 64% private) which related with special topic like technology, economics, social, bank, photography and so on.

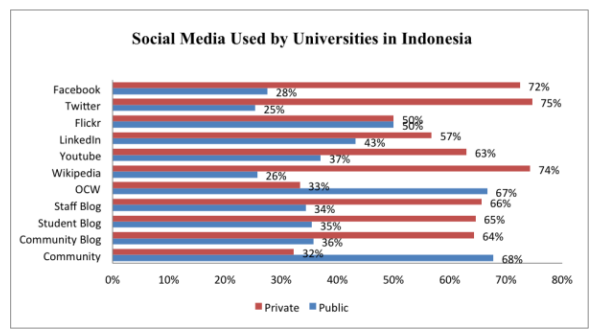

Figure 3.    Number of Social Media Used by Universities in Indonesia

Seventy-seven percent universities in Indonesia have own profiles on Wikipedia. Wikipedia is a social community and it takes the content that users submit very seriously. With 15 billion page views a month and a top position in 96.6 percent of search results.

Our research showed that 38 percent of universities have own lecturer blog, followed by 30 percent have own student blog. The best backlinks that can get is also from blog which the content is good and also with significant traffic.

Indonesia is the 4th-largest Facebook nation in the world and the 5th-largest Twitter users. Not all universities have official Facebook (26%) and Twitter (25%). However, these social media remain largely untapped on education to spread information about university.

LinkedIn is social media that huge with teachers and education professionals. According to recent release from the LinkedIn's blog, there are now more than 200 million members as part of LinkedIn network. Similarly with YouTube, a wide range of videos with educational value are available on YouTube. Everyone can watch full courses from the world's leading universities, professional development material from fellow educators, and inspiring videos from global thought leaders. Flickr is also exciting site to share photos online, create groups that are public or entirely private. Flickr is to be a great tool to easily share photos with students, alumni, faculty and staff. The automation of uploading photos and turning them into organized collections with slideshows is a great timesaver for time-strapped web development staff. OCW is Open Course Ware, originally launched by the Massachusetts Institute of Technology (MIT) in 2001. The stated goal of the program is to provide the content that supports an education [27]. This is a great resource for self-improvement and for college students who would like extra guidance through courses at their own institutions. Many students like the reinforcement of studying lecture notes and materials from parallel courses. The content is amazingly rich and can include online textbooks, exams, images and sometimes video or audio clips. Unfortunately, few universities in Indonesia joined LinkedIn (14%), Youtube (10%) and Flickr (3%) as social media platform presence. OCW is the least used as social media platform among universities in Indonesia (1%). Actually, sharing educational resources over the Internet provides multiple benefits, from academic collaboration to economic development.

Alexa traffic ranking is a website indexing engine that attempts to measure popularity of websites. The lower the Alexa ranking, the more popular of website is viewed. In Table 2 we can see that the average global ranking of Universities in Indonesia which have social media are 1.181.152. The rank followed by country average on 14.679 and 2.561 for number of reputation link. The following are explanation about the rank elements in Alexa:

- *Global*, an estimate of website's popularity. The rank is calculated using a combination of average daily visitors to a website and pageviews on a website over the past 3 months. The site with the highest combination of visitors and pageviews is ranked #1.

- *ID*, an estimate of website's popularity in a specific country (Indonesia). The rank by country is calculated using a combination of average daily visitors to a website and pageviews on a website from users from that country over the past month. The site with the highest combination of visitors and pageviews is ranked #1 in that country (Indonesia).

- *Reputation Link*, a measure of a website's reputation. The number of links to a website from sites visited by users in the Alexa traffic panel. Links that were not seen by users in the Alexa traffic panel are not counted. Multiple links from the same site are only counted once.

TABLE II. AVERAGE ALEXA RANK (HAVE/HAVEN'T SOCIAL MEDIA)

| Alexa | | |
|---|---|---|
| **Have Social Media** | *Global* | *ID* | *Reputation Link* |
| | 1.181.152 | 14.679 | 2.561 |
| **Have not Social Media** | *Global* | *ID* | *Reputation Link* |
| | 5.091.057 | 21.018 | 397 |

Majestic SEO is a link intelligence tools for SEO and internet page rank and also marketing. It is used for Webometrics research to evaluate the impact indicator. The quality of the contents is evaluated through a "virtual referendum", counting all the external inlinks that the University webdomain receives from third parties. Those links are recognizing the institutional prestige, the academic performance, the value of the information, and the usefulness of the services as introduced in the webpages according to the criteria of millions of web editors from all over the world. As it is seen from the table below (Table 3), universities which have official social media have on average 1.726 for referring domain, 231.377 for external backlinks, 31 for citation flow and 28 for trust flow. There are most significant differences for result of popularity by Alexa Rank and visibility by Majestic SEO in universities whether use social media or no. The following are explanation about Majestic SEO elements:

- *Reffering Domain*, also known as "ref domain", is a domain from which a backlink is pointing to a page or link.

- *External Backlinks*, also referred to in SEO as a external "inlink".

- *Citation Flow*, is a Majestic SEO Flow Metric, which is weighted by the number citations to a given URL, or Domain.

- *Trust Flow*, is a Majestic SEO Flow Metric, which is weighted by the number of clicks from a seed set of trusted sites to a given URL, or Domain.

TABLE III.     AVERAGE MAJESTIC SEO (HAVE/HAVEN'T SOCIAL MEDIA)

| Majestic SEO | | | | |
|---|---|---|---|---|
| Have Social Media | *Referring Domain* | *External Backlinks* | *Citation Flow* | *Trust Flow* |
| | 1.726 | 231.377 | 31 | 28 |
| Have not Social Media | *Referring Domain* | *External Backlinks* | *Citation Flow* | *Trust Flow* |
| | 323 | 20.249 | 19 | 14 |

The difference in social media use among public and private universities examine with the chisquare with the following test results show below in Table 4.

TABLE IV.     CHISQUARE TEST OF SOCIAL MEDIA OWNERSHIP

| Type of Social Media | Pearson chisquare | Sign. |
|---|---|---|
| Lecturer's blog | 20.096 | 0.000 |
| Student's blog | 16.592 | 0.000 |
| Community blog | 0.665 | 0.415 |
| Facebook | 3.240 | 0.072 |
| Twitter | 1.570 | 0.210 |
| Flickr | 4.604 | 0.032 |
| LinkedIn | 14.394 | 0.000 |
| Youtube | 1.711 | 0.191 |
| Wikipedia | 17.214 | 0.000 |

The Chisquare test result indicate that there are significant differences between public and private universities in the utilization of lecturer's blog, student's blog, Flickr, LinkedIn, and Wikipedia. If it is seen from the value of Chi-Square, social media type of the highest difference is Lecturer's blog, followed by Wikipedia, Student's blogs, and LinkedIn. Four types of other social media showed no difference between public and private universities, that are community blog, facebook, Twitter, and Youtube. If we look at the types of social media that have different significant, it can be concluded that public universities tend to use more social media that can gather posts of lecturers and students in the form of individual blogs.

One of purpose of the use of social media is dissemination of information to the community, particularly the academic community of each college. If the utilization of social media is effective, the number of visits to the college website will increase, or in other words, will increase the popularity of the website. Differences the popularity of the website between college based on ownership of social media tested by independent samples t test with complete results can be seen in the Table 5 below.

TABLE V.     RESULT OF INDEPENDENT T TEST

| Type of SocMed | Signifinces of web popularity[a] | | | | | |
|---|---|---|---|---|---|---|
| | RF | TL | CF | TF | GR | RL |
| Lecturer'blog | Yes | Yes | Yes | Yes | Yes | Yes |
| Student's blog | Yes | Yes | Yes | Yes | Yes | Yes |
| Community blog | No | No | Yes | Yes | Yes | Yes |
| Facebook | Yes | Yes | Yes | Yes | Yes | Yes |
| Twitter | Yes | No | Yes | Yes | Yes | Yes |
| Flickr | Yes | Yes | Yes | Yes | Yes | Yes |
| LinkedIn | Yes | Yes | Yes | Yes | Yes | Yes |
| Youtube | Yes | Yes | Yes | Yes | Yes | Yes |
| Wikipedia | Yes | No | Yes | Yes | Yes | Yes |

[a] RF=Reffering domain, TL=Total backlink, CF=Citation Flow, GR=Global Traffic Rank, RL=Reputation Link

The result of independent t test showed that the overall ownership of social media has a significant impact on the popularity of the website. Negative results occur only in the community blog toward reffering domain and total backlinks, as well as Twitter and Wikipedia toward total backlinks.

## V.     CONCLUSION

The study concluded that universities in Indonesia used social media with more interest in Facebook as it became most popular amongst others, then followed by Twitter. Seventy-seven percent universities in Indonesia have their own profiles university on Wikipedia. Our research also showed that 38 percent of universities have own lecturer blog, followed by 30 percent have own student blog. Unfortunately, few universities in Indonesia joined LinkedIn (14%), Youtube (10%) and Flickr (3%) as social media platform presence. Open Course Ware is the least used as social media platform among universities in Indonesia (1%).

The average Alexa global ranking of Universities in Indonesia which have social media are 1.323.912. The rank followed by country (ID) average on 15.581 and 2.511 for number of reputation link. On Majestic SEO, universities which have official social media average on 1.686 for referring domain, 235.153 for external backlinks, 31 for citation flow and 28 for trust flow. There are most significant differences for result of popularity by Alexa Rank and visibility by Majestic SEO in universities whether use social media or no.

Public universities tend to use more different types of social media than private with differences test that significant for five types of social media, the lecturer's blog, student's blog, Flickr, LinkedIn, and Wikipedia. Use of social media has a significant impact to the popularity of the website. The analysis result showed the colleges that use social media have level of the popularity of the website is higher for all parameters of the popularity of the website are reffering domain, total backlinks, citation flow, trust flow, global traffic rank and reputation link. The exception applies only to the community blog toward reffering domain and total backlinks as well as Twitter and Wikipedia toward total backlinks.

## REFERENCES

[1] Internet Usage World Stats - Internet and Population Statistics. www.internetworldstats.com. 2012.

[2] Daily Social. MarkPlus Insight Survey: Indonesia Has 55 million Internet Users, 2011, Retrieved 26 December 2012 from http://dailysocial.net/en/2011/11/01/markplus-insight-survey-indonesia-has-55-million-internet-users/

[3] A. Farzaneha, K. Payam, O. Zahra and Abbas A. K. "Webometrics analysis of Iranian University of Medical Sciences," Scientometrics, 80(1), 255-266, 2009.

[4] I. Hussain and N. Gulrez. "Academic Use of Social Media: Practices and Problems of University Students," International Conference on Education and Management Innovation, 2012, IACSIT Press, Singapore. Retrieved 27 December 2012 from www.ipedr.com/vol30/37-ICEMI%202012-M00074.pdf

[5] S. Hazari, A. North and D. Moreland. "Investigating pedagogical value of wiki technology," Journal of Information Systems Education, vol. 20, pp. 187-198, 2009.

[6] T. M. Farwell and R. D. Waters. "Exploring the use of social bookmarking technology in education: an analysis of students' experiences using a course-specific Delicious.com account," Journal of Online Learning and Teaching, vol. 6, pp. 398-408, 2010.

[7] J. Armstrong and T. Franklin. "A review of current and developing international practice in the use of social networking (Web 2.0) in higher education," 2008. Retrieved 25 December 2012 from http://www.franklinconsulting.co.uk

[8] I. Hussain. "A study of emerging technologies and their impact on teaching learning process," An unpublished PhD Dissertation, Allama Iqbal Open University Islamabad, 2005.

[9] J. Armstrong and T. Franklin. "A review of current and developing international practice in the use of social networking (Web 2.0) in higher education," 2008. Retrieved 25 December 2012 from http://www.franklinconsulting.co.uk

[10] J. Armstrong and T. Franklin. "A review of current and developing international practice in the use of social networking (Web 2.0) in higher education," 2008.

[11] S. Hamid, S. Chang, and S. Kurnia. "Identifying the use of online social networking in higher education," Same places, different spaces. Proceedings Ascilite Auckland 2009. Retrieved 25 December 2012 from http://www.ascilite.org.au/conferences/auckland09/procs/hamid-poster.pdf

[12] N. Dabner. "Design to support distance teacher education communities: A case study of a student–student ementoring initiative," Proceedings of Society for Information Technology and Teacher Education International Conference 2011. Nashville, TN: AACE 1-880094-84-3., 2011.

[13] D. Mack, A. Behler, B. Roberts, and E. Rimland. "Reaching students with Facebook: Data and best practices," Electronic Journal of Academic and Special Librarianship, 2007, 8(2).

[14] N. E. Davis, N. Dabner, J. Mackey, D. Morrow, C. Astall, J. Cowan, and et al. "Converging offerings of teacher education in times of austerity: Transforming spaces, places and roles," Proceedings of Society for Information Technology and Teacher Education International Conference 2011. Nashville, TN: AACE 1-880094-84-3, 2011.

[15] M. D. Roblyer, M. McDaniel, M. Webb, J. Herman, and J. V Witty. "Findings on Facebook in higher education: A comparison of college faculty and student uses and perceptions of social networking sites," The Internet and Higher Education, 13 (3), 134–140, 2010.

[16] C. Madge, J. Meek, J. Wellens, and T. Hooley. "Facebook, social integration and informal learning at university: It is more for socializing and talking to friends about work than for actually doing work,". Learning, Media and Technology, 34(2), 141–155, 2009.

[17] G. Bull, , A. Thompson, , M. Searson, J. Garofalo, J. Park, J. Young and J. Lee. "Connecting informal and formal learning experiences in the age of participatory media," Contemporary Issues in Technology and Teacher Education, 2008, 8(2). Retrieved from http://www.citejournal.org/vol8/iss2/editorial/article1.cfm [accessed 25 December 2012].

[18] C. Y. Mason and R. Dodds. "Bridge the digital divide for educational equity," Education Digest, 70(9), 25-27, 2005a.

[19] C. Y. Mason and R. Dodds. "Bridging the digital divide: Schools must find ways to provide equal access to technology for all students," Principal, 84(4), 24-30, 2005b.

[20] K. Mossberger, C. J. Tolbert and R. McNeal. "Digital Citizenship: The Internet, Society, and Participation," Massachusetts: The MIT Press, 2008.

[21] V. Venkatesh and M. G. Morris. "Why Don't Men Ever Stop Ask for Direction? Gender, Social Influence and Their Role in Technology Acceptance and Usage Behavior," MIS Quarterly (24:1), pp 115-139, 2000.

[22] J. Olson, M. Clough and K. Penning. "Prospective elementary teachers gone wild? An analysis of Facebook selfportraits and expected dispositions of preserve elementary teachers," Contemporary Issues in Technology and Teacher Education, 9(4), 443–475, 2009.

[23] J. Grimmelmann. "Saving Facebook," Iowa Law Review, 94, 1137–1206, 2009. Retrieved 26 December 2012 from http://works.bepress.com/cgi/viewcontent.cgi?article=1019&context=james_grimmelmann

[24] J. Cain, D. Scott, and P. Akers. "Pharmacy students' Facebook activity and opinions regarding accountability and eprofessionalism," American Journal of Pharmaceutical Education, 2009, 73, 6. Retrieved 26 December 2012 from http://www.ajpe.org/aj7306/aj7306104/aj7306104.pdf

[25] A. Hewitt and A. Forte. "Crossing boundaries: Identity management and student/faculty relationships of Facebook," Presented at CSCW06, November 4–8, Alberta, Canada, 2006.

[26] N. Selwyn. "Face working: Exploring students' education-related use of Facebook," Learning, Media and Technology, 34(2), 157–174. 2009.

[27] K. Kirkpatrick. "Opencourseware: an MIT thing?," Searcher. 14(6). pp 53-58, 2006.

[28] I. Ahmed and T. F. Qazi. "A look out for academic impacts of Social networking sites (SNSs): A student based perspective," African Journal of Business Management, vol. 5(12), pp. 5022-5031, ISSN 1993-8233, 2011.

[29] P. Rambe. "Exploring the Impacts of Social Networking Sites on Academic Relations in the University," Journal of Information Technology Education, vol. 10, 2011.

[30] N. Eteokleous. "Facebook - A Social Networking Tool For Educational Purposes: Developing Special Interest Group," ICICTE Proceedings, 2012.

## AUTHORS PROFILE

Hanum Putri Permatasari is a lecturer from Department of Information System faculty of Computer Science and Information Technology in Gunadarma Universtiy. She received her MMSI of Information System from Gunadarma University. Website development in higher education is one of her interest research.

Silvia Harlena is a lecturer from Department of Information System faculty of Computer Science and Information Technology in Gunadarma Universtiy. She received her S.KOM of Information System from Gunadarma University. Website development in higher education is one of her interest research.

Donny Erlangga is a lecturer from Department of Informatics Engineering Diploma Program in Gunadarma Universtiy. He received his MMSI of Information System from Gunadarma University. Computer Network is one of his interest research.

Reza Chandra is a lecturer from Department of Information System faculty of Computer Science and Information Technology in Gunadarma Universtiy. He received his MMSI of Information System from Gunadarma University. Computer Network is one of his interest research.